\documentclass[a4paper]{article}
\usepackage{natbib}
\usepackage{graphicx}
\usepackage{amsmath}
\usepackage[margin=2.5cm]{geometry}

\begin{document}

%
%


\title{Coherent changes of the circulation in the deep North Atlantic from moored transport arrays}

%
%




\author{E. Frajka-Williams\footnote{Ocean and Earth Science, University of Southampton, Southampton, SO14 3ZH}, M. Lankhorst, J. Koelling, and U. Send\footnote{Scripps Institution of Oceanography, University of California San Diego, La Jolla, USA}}









\maketitle

\begin{itemize}
\item Southward flow weakened below 3 km relative to above by 2.6--3.9 Sv.
\item The shift occurred between 2009--2010, and earlier at $26^\circ$N than $16^\circ$N. 
\item From $26^\circ$N observations, geostrophic reference level methods can influence transport trends. 
\end{itemize}

%
%


\begin{abstract}
In situ boundary arrays have been installed in the North Atlantic to measure the large-scale ocean circulation. Here, we use measurements at the western edge of the North Atlantic at $16^\circ$N and $26^\circ$N to investigate  low-frequency variations in deep densities and their associated influence on ocean transports. At both latitudes, deep waters (below 1100 dbar) at the western boundary are becoming fresher and less dense. The associated change in geopotential thickness is about $0.15$ $\mbox{m}^2\mbox{s}^{-2}$ between 2004--2009 and 2010--2014, with the shift occurring between 2009--2010 and earlier at $26^\circ$N than $16^\circ$N.   Without a similar density change on the east of the Atlantic, a mid-depth reduction in water density at the west drives an increase in the shear between the upper and lower layers of North Atlantic Deep Water of about 2.6 Sv at $26^\circ$N and 3.9 Sv at $16^\circ$N.  While these transport anomalies result in an intensifying tendency in the meridional overturning circulation (MOC) estimate at $16^\circ$N, the method of applying a zero net mass transport constraint at $26^\circ$N results in an opposing (reducing) tendency of the MOC.
\end{abstract}

%
%

%


%
%
%
%

\section{Introduction}
The large-scale ocean circulation is often displayed in schematics with ribbons of red and blue indicating warm and cold transports at different depths. These schematics capture several key aspects of the meridional overturning circulation (MOC): that it includes  warm thermocline waters flowing northwards in the top 1000 m of the Atlantic and colder waters at depth moving generally southwards.  The thermocline waters carry heat northwards, while the deep waters, recently formed through interaction with the atmosphere at the surface, store carbon and other properties at depth. Zonally-averaging this circulation across the Atlantic basin from east-to-west, the meridional flow (flow in the north-south direction) shows ``overturning'' with surface waters moving northwards, deepening, then returning southwards at depth \citep{Danabasoglu-etal-2014}.  The strength of the overturning then refers to the total northward flow in the top $\sim$1000 m of the Atlantic, which is equal and opposite to the southward flow below.  This overturning is typically about 17~Sv (1 Sv = 1,000,000 $\mbox{m}^3\mbox{s}^{-1}$).

Schematics of overturning, while capturing some of the salient features, also connote a circulation that is simple and laminar, and when referred to as a ``Great conveyor'' suggest a conveyor belt moving at similar speeds everywhere.    While time-mean circulation shows a continuous northward flow across the tropics to mid-latitudes in the Atlantic, variations in the strength of overturning at different latitudes may not be simultaneous.  A long simulation (1000 years) of the time-varying overturning circulation identified lower frequency fluctuations in the subpolar regions and interannual variations in subtropical regions \citep{Zhang-2010}. In particular, the subtropical transport magnitude exhibited variations of the same sign as those in the subpolar regions, but at some time delay.   More realistic simulations investigating the coherence of the overturning find that across the subtropics, fluctuations are relatively coherent, meaning instantaneously correlated, on interannual timescales ($r>0.6$ between $0$--$40^\circ$N) \citep{Bingham-etal-2007}.  Differences in the strength of overturning between latitudes may result in local convergences or divergences of heat \citep{Cunningham-etal-2014,Kelly-etal-2014} which may in turn drive heat fluxes into or out of the atmosphere.

Moored estimates of the time-varying transports in the Atlantic show substantial interannual and sub-annual variability \citep{FrajkaWilliams-etal-2016,Send-etal-2011,Toole-etal-2011}.  However, efforts to link observations between distant individual latitudes have shown limited meridional coherence of the MOC \citep{Elipot-etal-2014,Mielke-etal-2013}.  On sub-annual time scales, fluctuations between latitudes appear to be out-of-phase.  \citet{Mielke-etal-2013} showed that the seasonal cycles of the non-Ekman component of the overturning were $180^\circ$ out-of-phase between $26^\circ$N and $41^\circ$N, though the phasing of the observed seasonal cycle at $41^\circ$N did not agree with the modeled seasonal cycle.  \citet{Elipot-etal-2014} also identified an out-of-phase relationship between the large-scale transport fluctuations at different latitudes but using only the western boundary density signals to compute transports.  Some of these fluctuations in transport have been found to have a fixed relationship to two modes of wind stress variability over the Atlantic \citep{Elipot-etal-2017},  with locations at $16^\circ$N and $26^\circ$N related to the first mode of variability, and the more northerly regions ($\sim40^\circ$N) to the North Atlantic Oscillation pattern of wind forcing.

On longer (interannual-to-one-decade) timescales, where the overturning circulation may be expected to represent larger-scale basin-wide fluctuations in ocean circulation, the MOC at $26^\circ$N has been shown to have a declining trend \citep{Smeed-etal-2014} while the transports estimated at $16^\circ$N \citep{Send-etal-2011} show a different tendency.  Transports at both latitudes are monitored using a boundary mooring approach, where temperature and salinity profiles are measured continuously at western and eastern edges, spanning great swaths of the ocean.  The method of calculating transports relies on the thermal wind relation between meridional shear in transports and zonal density gradients.  However, thermal wind only determines the velocity shear relative to a level of no or known motion.  The methods used to compute transports differ between the two latitudes in their application of a choice of reference level.    

Observed low frequency variation of transports at $16^\circ$N showed a weakening overturning from 2000--2010, resulting primarily from changes at the Mid-Atlantic Ridge \citep{Send-etal-2011}. From 2010 to 2014, transports are now strengthening, consistent with an intensification of the MOC $16^\circ$N.  This is in apparent contradiction with the transport fluctuations estimated at 26°N, where a reduction in the lower NADW layer (3000--5000 m) was identified over the 2004--2014 period \citep{Smeed-etal-2014,FrajkaWilliams-etal-2016}.    

In this paper, we explore whether or not the meridional overturning circulation is coherent (similar sign/magnitude changes) at $16^\circ$N and $26^\circ$N, and the influence of the method of calculation on the transport estimates.  In section 2, the data and methods are described. In section 3, we detail the hydrographic properties and tendencies over the 11 and 16 years of observations at the two latitudes.  In section 4, we use the hydrographic data to construct transport shear estimates from dynamic height, and discuss how the observed variations influence transbasin transport estimates.  Finally, in section 5 we conclude and highlight the key issue of the choice of reference level for transport estimates.

\section{Data \& Methods}
Data used here are from two mooring arrays in the Atlantic: the RAPID Climate Change (RAPID) and Meridional Overturning Circulation and Heat transport Array (MOCHA) moored observations at $26.5^\circ$N from 2004--2015 and the Meridional Overturning Variability Experiment (MOVE) moored observations at $16^\circ$N from 2004--2016 (Fig. 1a). Both arrays were designed to estimate the strength of the overturning circulation using boundary measurements (Fig. 1b).  

Two  major differences exist between the two arrays.  First, at  $16^\circ$N, the array extends eastward only to the Mid-Atlantic Ridge while at $26^\circ$N, it extends eastward to Africa.   For longer timescales (4+ years), observing system simulation experiments (OSSEs) identified that transport fluctuations are primarily due to density changes at the western boundary at $16^\circ$N \citep{Kanzow-2004}.  This is confirmed by observations at $26^\circ$N show that the western boundary dominates transport variability on interannual and longer timescales \citep{FrajkaWilliams-etal-2016}. There is some uncertainty resulting from ocean transports that may not be captured at $16^\circ$N east of the Mid Atlantic Ridge, and so we follow the method of \citet{Elipot-etal-2014} in focusing on the dynamic height variations at the western boundary only.   The second difference is in the application of a choice of reference level.  Traditionally, geostrophic shear is referenced to a level of no motion, where the integration (described further below) is referenced to a level where flow is weak or absent.  At $16^\circ$N, this is applied by choosing a deep reference level of no motion.  At $26^\circ$N, a deep reference level is chosen, but transports are then corrected with a barotropic velocity profile (as a hypsometric compesation term, $T_{ext}$) which is determined to ensure zero net mass transport across the latitude section.  This second difference, and its influence on the computed MOC transports, is explored in more detail in section \S4.2.

\begin{figure}[h]
\centering
\includegraphics[width=20pc]{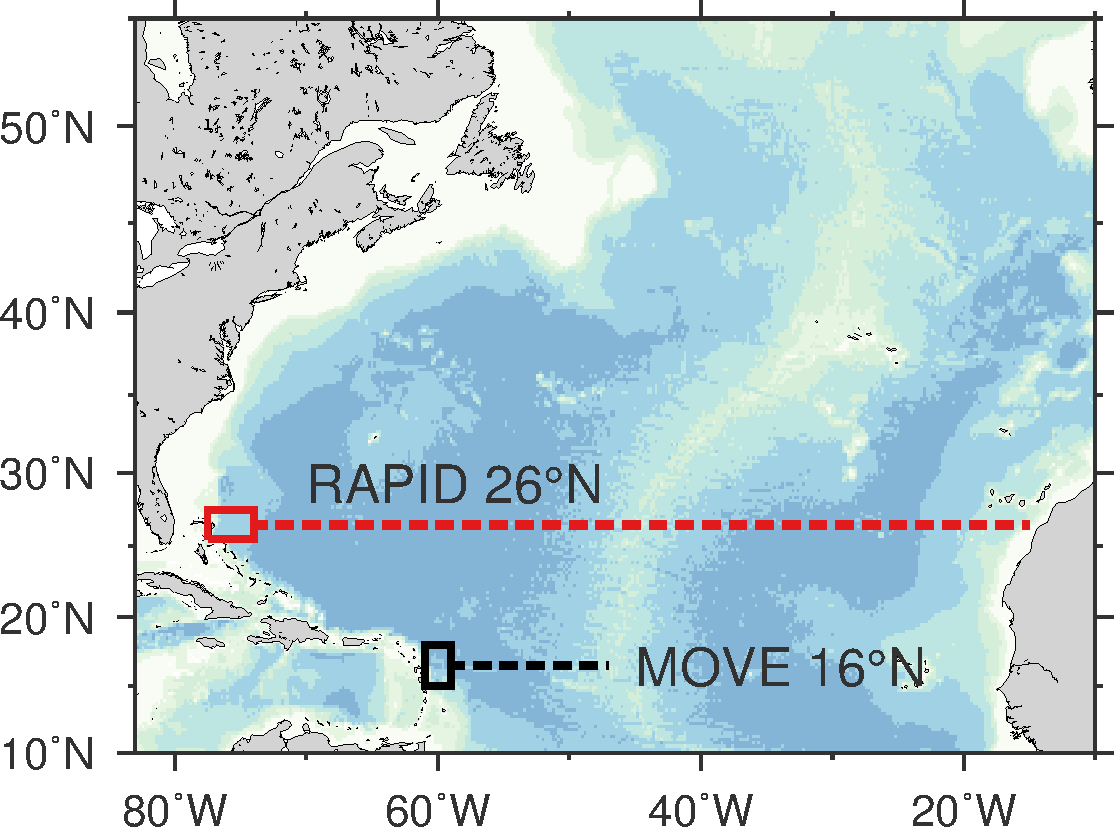}
\caption{Moored observations at RAPID 26$^\circ$N and MOVE 16$^\circ$N.  Bathymetry is shaded in color.}
\label{figone}
\end{figure}

\subsection{RAPID $26^\circ$N observations}
At the western boundary at $26^\circ$N, the primary dynamic height observations are from a full-depth mooring in 4000 m at $26.5^\circ$N, $76.75^\circ$W (WB2).  Below 4000 m, instrument records are taken from nearby (within 25 km) moorings.  Temperature and salinity records from individual instruments  are vertically interpolated  to form a western profile of hydrographic data as described in \citet{McCarthy-etal-2015}.  During November 2005 to March 2006, the WB2 mooring failed, and so during this period, data from the WB3 mooring at $26.5^\circ$N, $76.5^\circ$N were substituted. Typical instrument configurations on this mooring include  18 MicroCAT (Seabird Electronics, Bellevue, WA) records between 50 and 4800 dbar, though specific instrument locations and sampling intervals have varied over the 10 years of observations.  Field calibrations are carried out on individual instruments by mounting them to the conductivity-temperature-depth (CTD) rosette for pre-deployment and post-deployment casts. MicroCAT measurements are compared to those from the CTD at bottle stops, with drifts between the pre-deployment and post-deployment casts used to offset the time series observations. Individual instrument records are filtered with a 2-day low pass filter to remove the tides before gridding vertically to 20 m resolution. Full details of the data processing can be found in \citet{McCarthy-etal-2015}.

\subsection{MOVE $16^\circ$N observations}
At $16^\circ$N, data from the MOVE 1 and MOVE3 moorings are used here. MOVE3 is a single, sub-surface mooring that was initially deployed early 2000 and has been in operation ever since. The location is approximately $16.3^\circ$N, $60.5^\circ$W, at 5000 m water depth a short distance east of Guadeloupe, while MOVE1 is on the western flank of the Mid Atlantic Ridge at $51.5^\circ$W. Measurements of temperature, salinity, and currents are made from this platform \citep{Kanzow-etal-2006}. Instrumentation has varied over the years; the present configuration has 21 MicroCAT instruments for temperature and salinity covering the depth range from 50 m to the seafloor. Earlier deployments only covered the deeper layers below 1000 m.

Removal of sensor drift is performed as for the $26^\circ$N array, using CTD casts before deployment and after recovery \citep{Kanzow-etal-2006}. The calibrated, quality-controlled data are made publicly available through the OceanSITES data portals (www.oceansites.org). Data available at OceanSITES also includes six additional sites where MOVE has made observations, two of which are still in operation.

\subsection{Time series processing}
Data from both $26^\circ$N and $16^\circ$N were bin-averaged into monthly time series.  In order to focus on interannual and longer-term variations, a seasonal climatology was removed and time series were filtered with an 8-month Tukey window.  While some sub-annual variations remain, the $<1$-year filter window permits better identification of the timing of changes. In calculating correlations between time series, statistical significance was based on two-tailed t-tests where the numbers of degrees of freedom were determined from the integral time scale of decorrelation (Emery and Thomson, 2004).

\section{Hydrographic changes at $16^\circ$N and $26^\circ$N}
Temperature-salinity (TS) diagrams of water mass properties at $26^\circ$N and $16^\circ$N show variations from warm and salty in the thermocline to cold and fresh at depth, with only a modest change in slope of the T-S relationship around 2000 m (Fig. 2). This bend in the curve corresponds to the transition between central Labrador Sea Water (cLSW) and Iceland-Scotland overflow water (ISOW, \citet{Sebille-etal-2011}). At $26^\circ$N, in the recent 10 years, the waters below 1100 m have tended towards cold and/or fresh on an isopycnal with the exception of at 2000 m (cLSW) where the properties have remained the same. At $16^\circ$N, properties at all depths have tended towards cold and fresh on an isopycnal. These changes are consistent with, but smaller amplitude than, the cooling and freshening observed at $26^\circ$N from hydrographic sections from 1984--2010 \citep{Sebille-etal-2011}.

\begin{figure}[h]
\centering
\includegraphics[width=30pc]{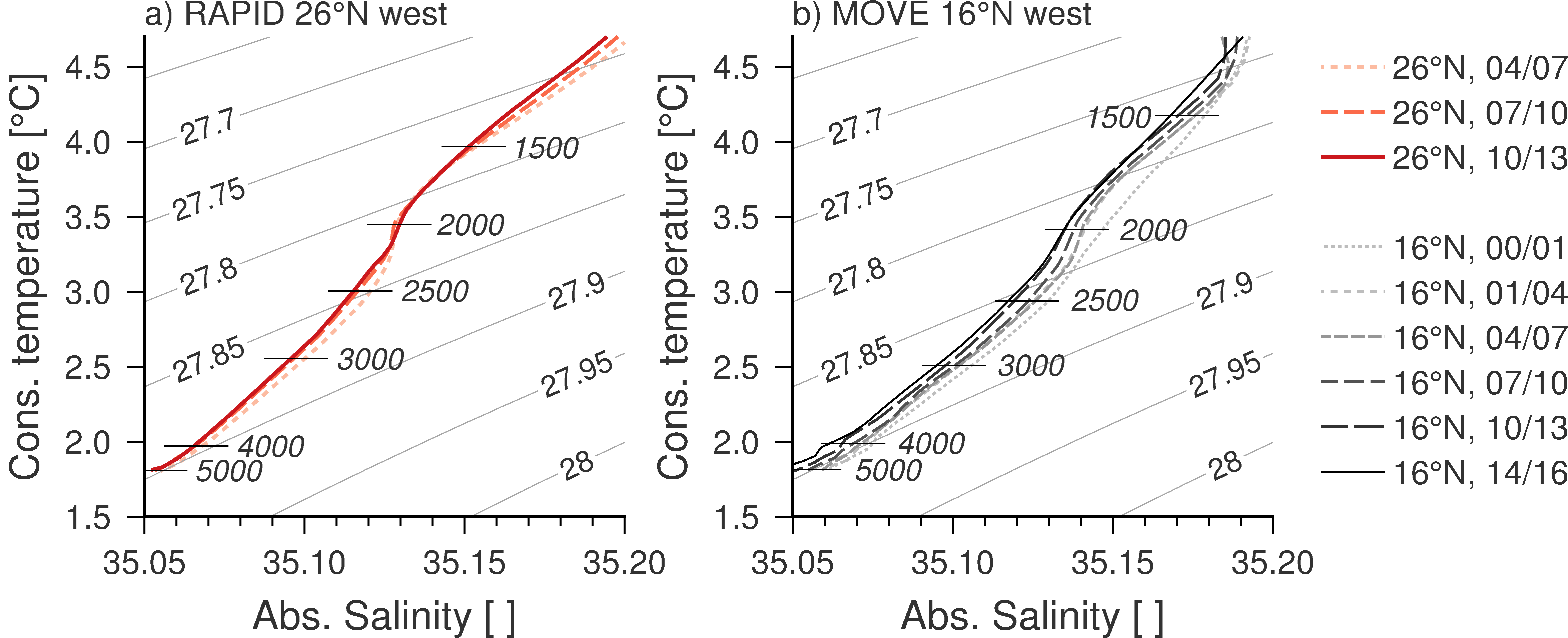}
\caption{3-year averages of the monthly binned conservative temperature and absolute salinity, between Oct 2000 and Oct 2015, where 00/03 indicates the period Oct 2000 through Sep 2003.  For the RAPID array at $26^\circ$N, the first averaging period $04/06^*$ represents the shorter period Apr 2004 through Sep 2006.  Average depths are indicated by the black lines.}
\label{figtwo}
\end{figure}

\subsection{Temperature and salinity changes on depth surfaces}
Both the $16^\circ$N and$26^\circ$N arrays are designed to capture transport variability. From this perspective, density changes on depth surfaces influence transport more than property changes on isopycnals. Temperatures at both latitudes decrease to a minimum at depth of about $1.8^\circ$C at $26^\circ$N and $1.9^\circ$C at $16^\circ$N (Fig. 3). Salinities at both latitudes are fresher at depth than in the thermocline. At mid-depths, warmer, saltier waters are found (around $3^\circ$C and $35.1$ around 2000 m). Over the 11-year RAPID deployment and 16-year MOVE deployment, the waters at depth (below 1000 m) have tended towards fresher water, though the temperature changes on depth surfaces are more ambiguous.

\begin{figure}[h]
\centering
\includegraphics[width=30pc]{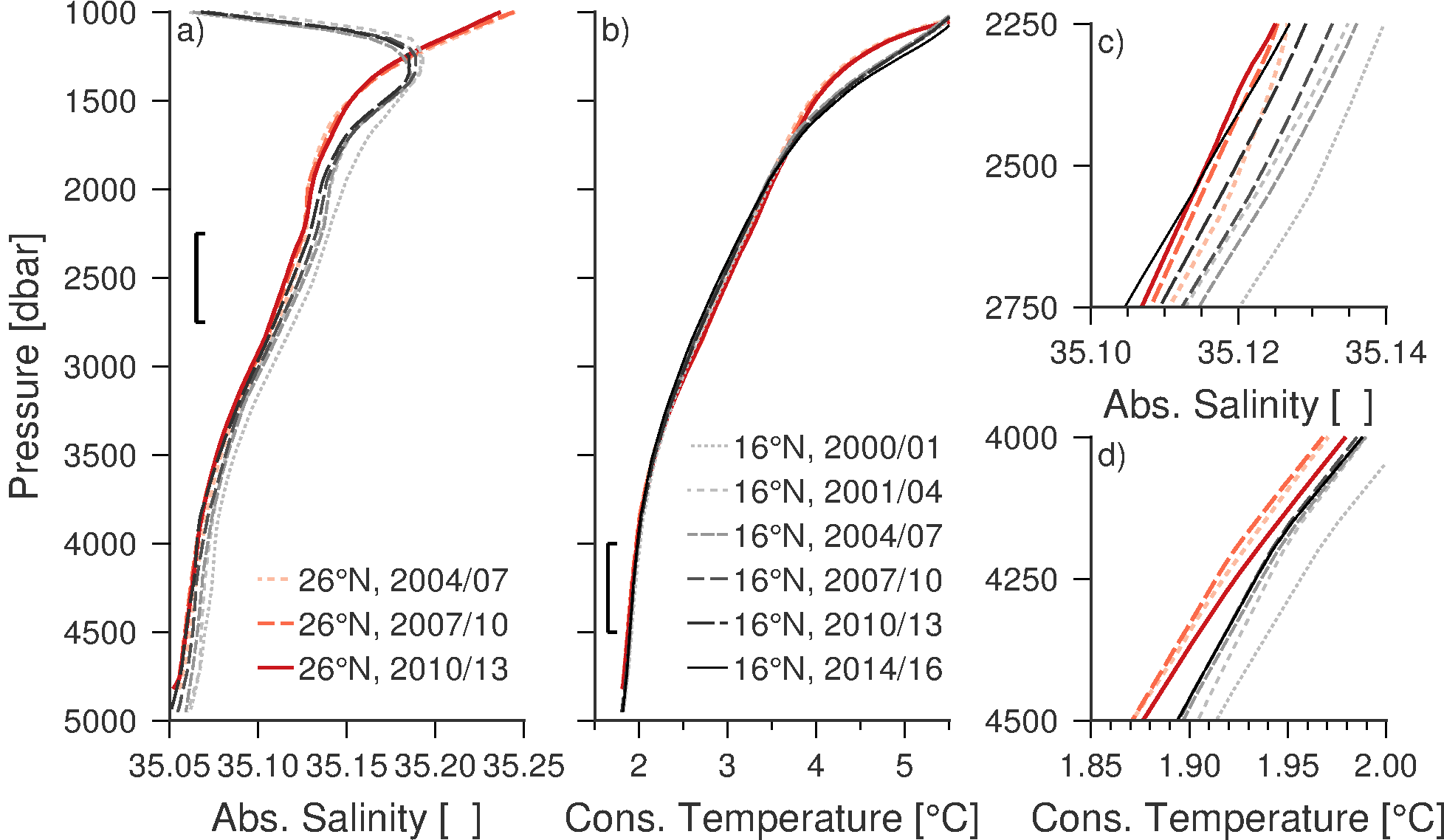}
\caption{(a) Absolute salinity and (b) conservative temperature  profiles from the western boundary of the RAPID 26$^\circ$N and MOVE 16$^\circ$N, in red and blue tones, respectively.   Darker colours indicate later 3-year averages as in Fig.~\ref{figtwo}.  (c) and (d) are insets of salinity and temperature for the depth ranges 2250--2750 and 4000--4500 dbar, respectively. }
\label{figthree}
\end{figure}

Property changes can be better seen in depth-time diagrams of the anomalies from the time-mean profile (Fig. 4--7). Over the 11-years of continuous RAPID deployments, temperatures below 2000 m have shifted from generally cooler to warmer, though the changes are neither monotonic nor depth-independent. In contrast, salinity changes are more monotonic on this 8-month smoothed data, transitioning from relatively salty to relatively fresh at all depths below 2000 m. At  $16^\circ$N at the west, temperature anomalies between 1500 and 3500 m show some warming from 2002--2013 (Fig. 5), while salinities have moved more monotonically relatively salty in 2000 to relatively fresh in 2015. The transition here does have a temporary reversal of the freshening tendency visible at mid-depth (1500--4000 dbar) during the 2009--10 winter, possibly associated with isopycnal heave.

\begin{figure}[h]
\centering
\includegraphics[width=20pc]{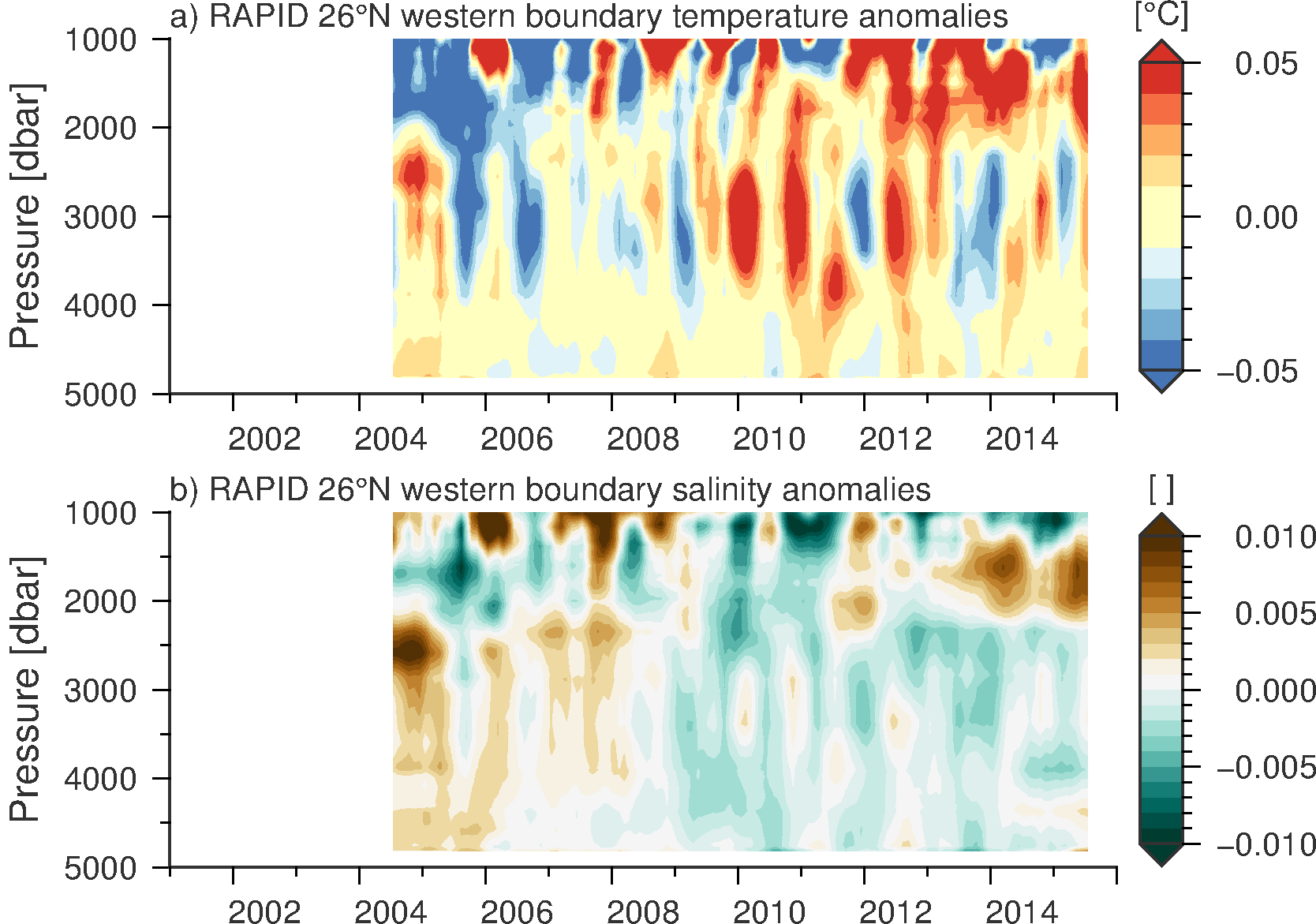}
\caption{Property anomalies at the western boundary composite profile from RAPID $26^\circ$N (around $26.5^\circ$N, $76.75^\circ$W). (a) Temperature anomalies are calculated relative to the mean profile over 2004-2014, with somewhat cooler temperatures prior to 2009. (b) Salinity anomalies relative to the mean profile over 2004--2014, with relatively fresher anomalies since 2008.   Properties at each depth have been smoothed with an 8-month Tukey filter.}
\label{figfour}
\end{figure}

\begin{figure}[h]
\centering
\includegraphics[width=20pc]{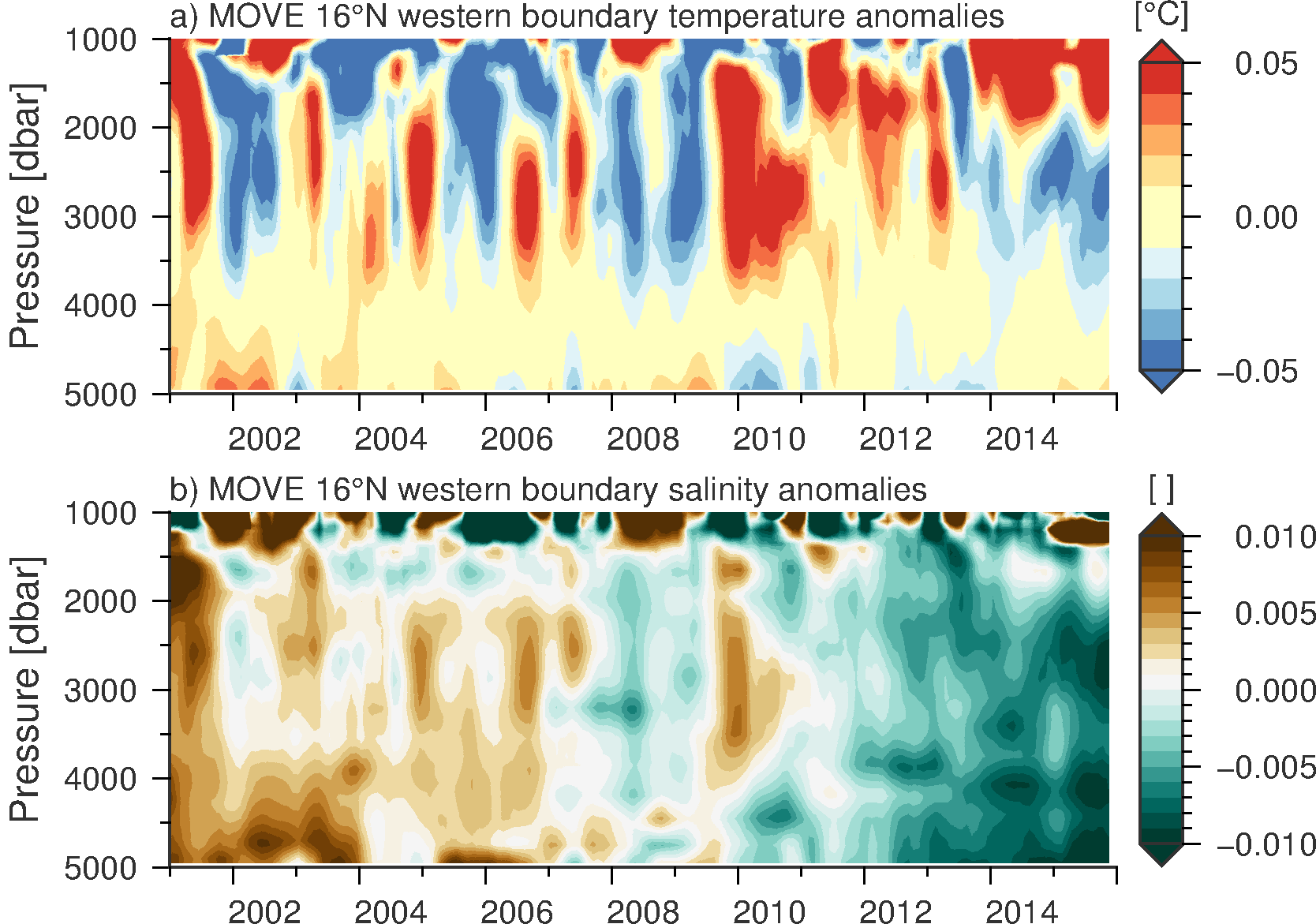}
\caption{As for Fig. 4, but for the western boundary of MOVE $16^\circ$N (mooring MOVE3, around $16^\circ$N, $60^\circ$W).}
\label{figfive}
\end{figure}

\begin{figure}[h]
\centering
\includegraphics[width=20pc]{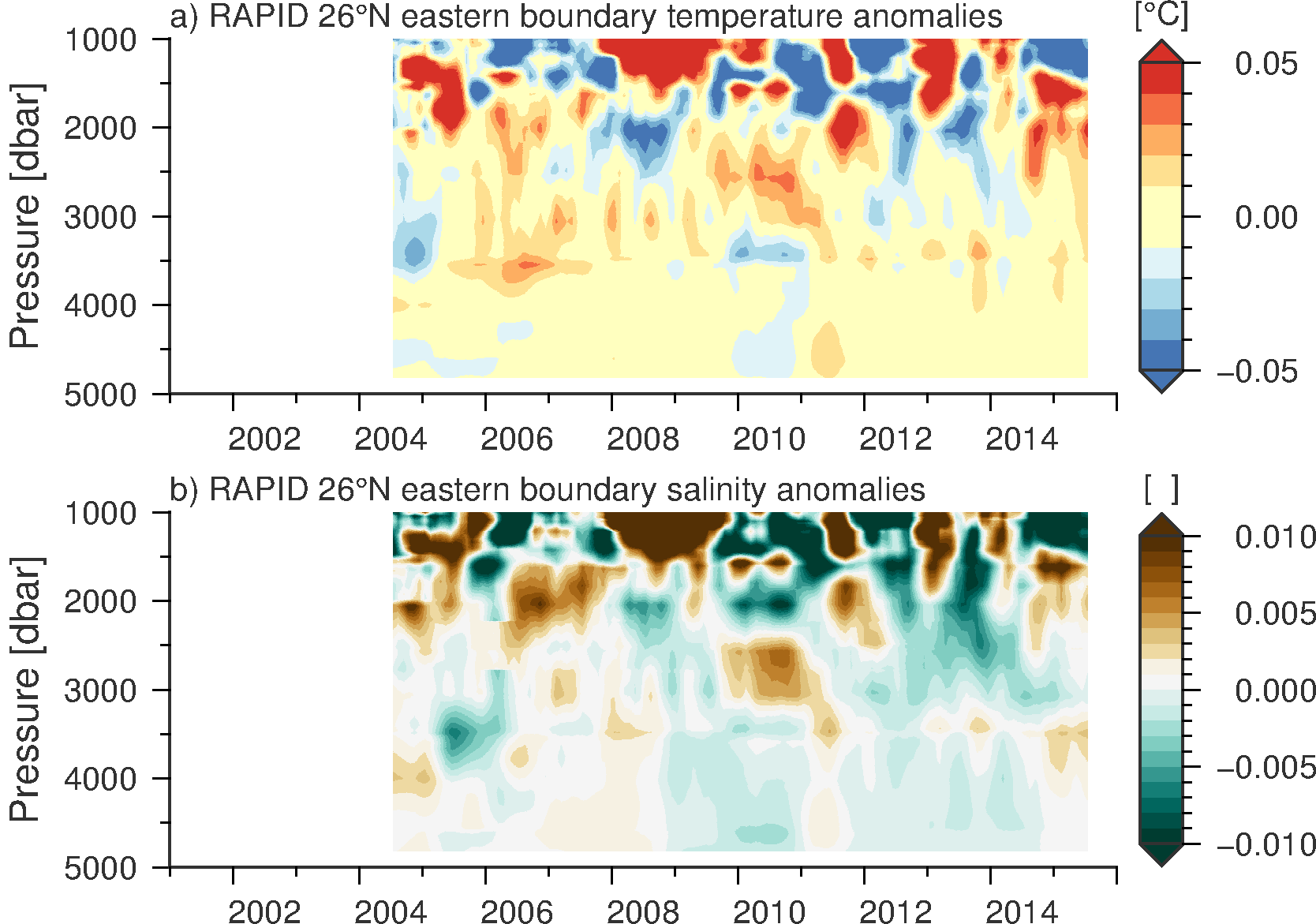}
\caption{As for Fig. 4, but for the eastern boundary composite profile from RAPID $26^\circ$N, east of the Mid-Atlantic Ridge up to the Canary Islands.}
\label{figsix}
\end{figure}

\begin{figure}[h]
\centering
\includegraphics[width=20pc]{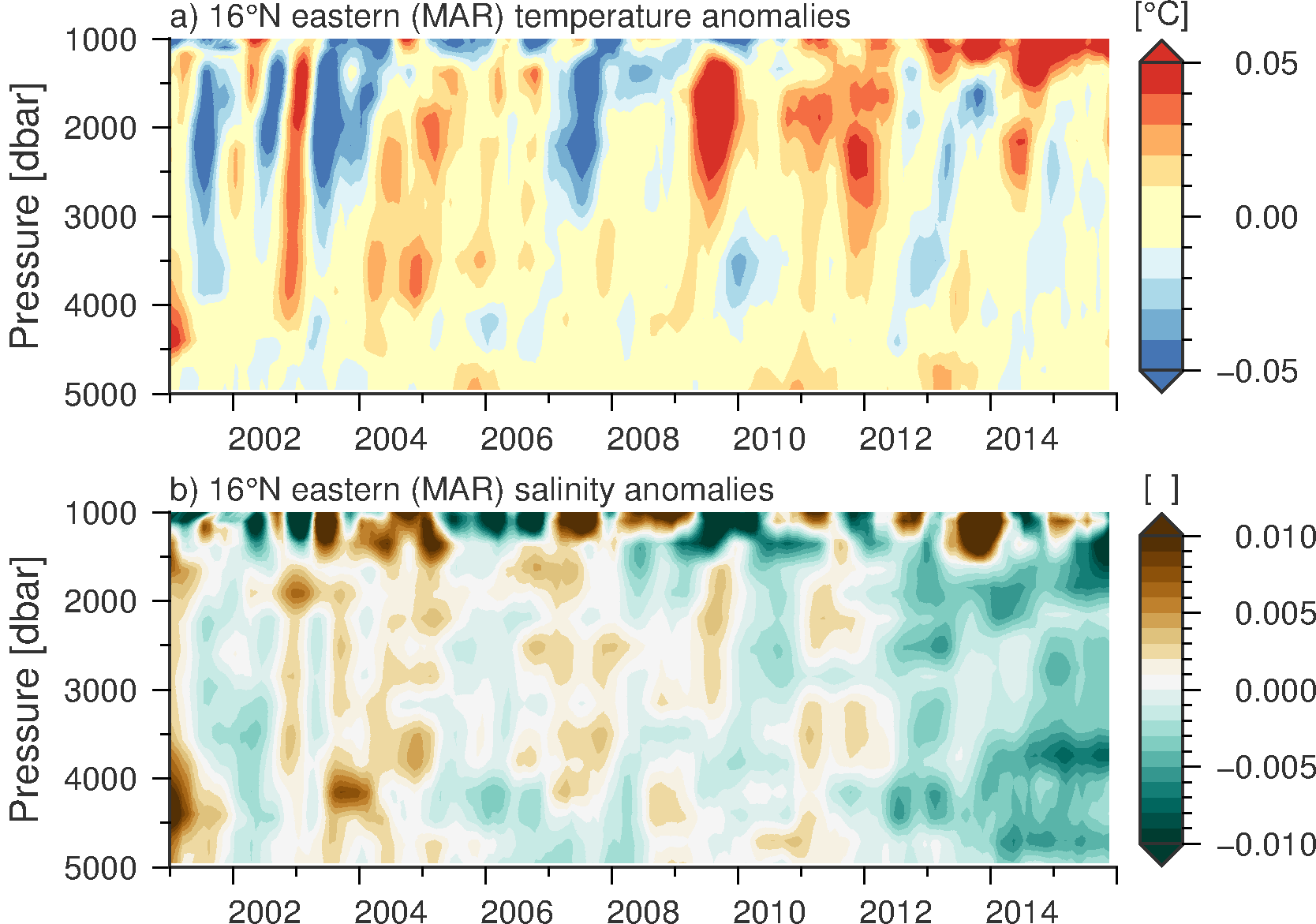}
\caption{As for Fig. 4, but for the eastern mooring from MOVE $16^\circ$N (MOVE1, around $15.5^\circ$N, $51.5^\circ$W).}
\label{figseven}
\end{figure}

Comparing the average properties between 1200 and 4650 dbar between the two complete five-year periods of observations, Apr 2004--Mar 2009 and Apr 2009--Mar 2014, temperatures warmed at $26^\circ$N by $0.023^\circ$C, from $2.77\pm0.03$ to $2.79\pm0.03^\circ$C, while salinities freshened by $0.002$, from $35.104\pm0.002$ to $35.103\pm0.002$. (Here, standard deviations are calculated on the annual means.) At $16^\circ$N, over the same two five-year periods, temperatures between 1200--4650 dbar warmed by $0.018^\circ$C, from $2.81\pm0.03$ to $2.83\pm0.03^\circ$C. Salinities freshened by about $0.003$, from $35.111\pm0.002$ to $35.108\pm0.003$. At both latitudes, the freshening results in lighter (less dense) waters at the western boundary at depth. While observed changes are near the estimated accuracy of measurements \citep{McCarthy-etal-2015}, the shift in properties between the two periods is statistically significant. We will see below that this change contributes to a shift in the estimated transport anomalies.

For completeness, we also show the property changes at the east (Fig. 6 and 7). In both cases, below 1200 dbar, anomalies are less coherent in time, with no apparent trend in temperature or salinity. This is consistent with the previous estimates and simulations indicating a reduced role for variations at the eastern boundary in controlling low-frequency transport fluctuations \citep{Kanzow-etal-2008,FrajkaWilliams-2015}.

Overall, property changes at depths below 2000 m show a tendency towards freshening and lighter water at both $16^\circ$N and $26^\circ$N.  These  changes cannot be easily used to distinguish whether the changes are due to  watermass changes (a different class of Labrador Sea Water)  or a vertical shift or heave of density surfaces with the same TS properties. The freshening is clear in TS space (towards freshening, Fig.~\ref{figtwo}), though whether it explains the density anomaly exclusively, or vertical heave is important, is left for future investigation.  Initial attempts to  separately diagnose property changes on density surfaces proved complicated, due to  small changes in salinity or temperature which affect both the calculation of the density surface and the property on that surface.

\subsection{Density changes}

Density anomalies rather than property changes can directly result in changes in circulation.  A density change on the western boundary of a basin, without compensating changes at the eastern boundary, will change the slope of the isopycnals across the basin. Through the thermal wind relation, a change in the zonal slope of  isopycnals  contributes to vertical shear in the meridional velocities. This is the fundamental principle behind the design of both observing arrays at $16^\circ$N and $26^\circ$N. Here, we investigate the density changes associated with aforementioned property changes at the western boundary of the Atlantic.

We first investigate the vertical coherence of density changes at a single latitude. At each depth, we compute a time series of density anomalies from the time mean. Time series are then correlated with each other to identify covariability between density anomalies at different depths. Fig. 8 shows the correlation coefficient between density anomalies at one depth (x-axis) with those at another depth (y-axis). Since density anomalies at the same depth are exactly correlated (correlation coefficient of 1), the 1-1 axis from upper left to lower right is exactly 1. Broader patches of high correlation around this axis are found for $26^\circ$N in the depth range 1100--4800 m and for $16^\circ$N in the range 1200--4650 dbar. This means, for example, that density anomalies at 2000 m co-vary with density anomalies at 4000 m. Overall, it suggests that density anomalies everywhere below 1200 m co-vary, or similarly, that the time series of density anomalies at a single depth will represent the variability for the whole deep layer. In contrast, in the thermocline above 1200 m, the red areas of co-variability contract back together towards the 1-1 diagonal (Fig. 8), indicating that density anomalies above 1200 m and do not co-vary with density anomalies below 1200 m.

\begin{figure}[h]
\centering
\includegraphics[width=30pc]{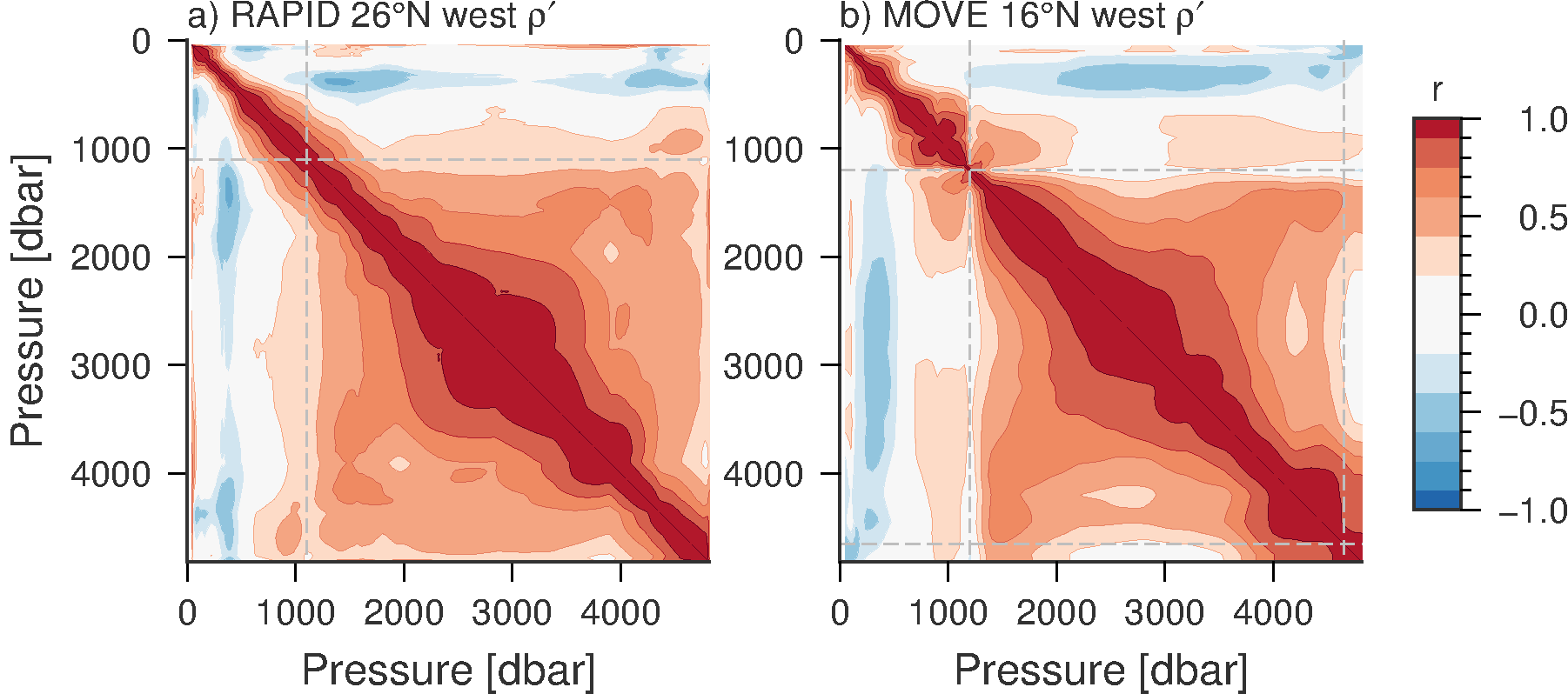}
\caption{Correlation between density anomalies at each depth from (a) the western boundary of RAPID $26^\circ$N, and (b) the western boundary of MOVE $16^\circ$N.  Red colours indicate positive correlation (coherent variations) while blue colours indicate negative correlation (anti-phase variations).}
\label{figeight}
\end{figure}

The high degree of covariability below 1200 m simplifies a comparison between latitudes because the density fluctuations everywhere below 1200 m will not depend strongly on the particular choice of depth.  It also suggests that a layered approximation of the ocean may be appropriate in the subtropical North Atlantic, with a small number of layers explaining a large fraction of the observed density variations.

\subsection{Timing of density changes}
While the time series of observations are relatively short for investigating interannual variations, and data have been smoothed with an 8-month window, we briefly investigate the relative timing of changes at the two latitudes. Given the vertical coherence of density variations at individual latitudes, we now compare the timing of density fluctuations between latitudes at the same depths. From visual inspection of the depth-time plots of property anomalies (Fig. 4 and 5), properties at $26^\circ$N shifted around 2009, while the shift at $16^\circ$N    either occurred around 2008 or 2010. Using density anomalies at each depth and latitude, the time series of density anomalies at, e.g., 2000 m at $26^\circ$N can be lag-correlated with the time series of density anomalies at 2000 m at $16^\circ$N. This lag correlation can be used to identify whether fluctuations in density at $26^\circ$N typically occur before or after fluctuations at $16^\circ$N.

\begin{figure}[h]
\centering
\includegraphics[width=20pc]{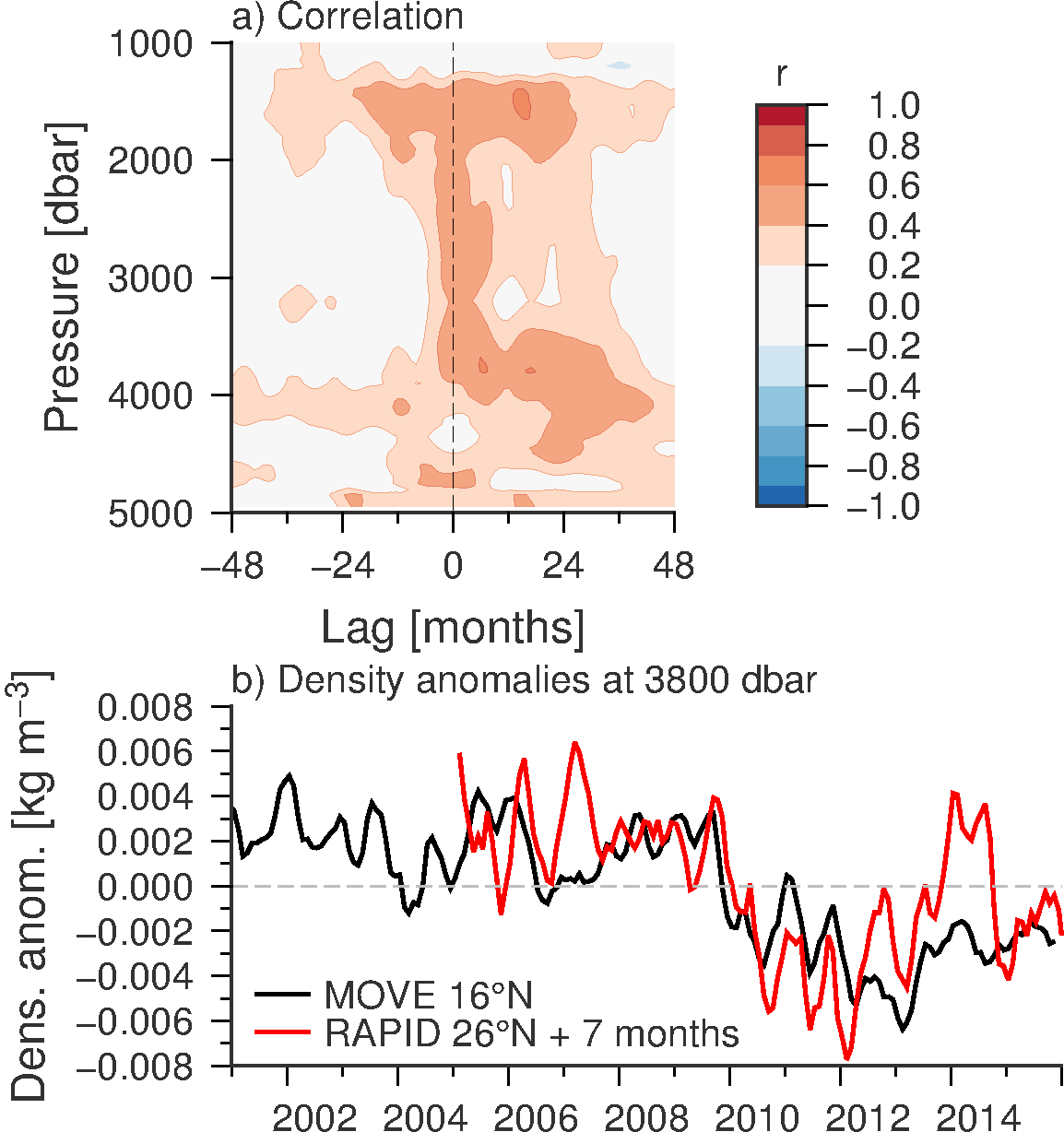}
\caption{Lag correlation between density anomalies at different latitudes but the same depth. (a) Correlation coefficient between density anomalies at the western boundaries of MOVE $16^\circ$N and RAPID $26^\circ$N, as a function of depth (y-axis) and lag in months (x-axis). (b) Time series of density anomalies at the two latitudes, at 3820 dbar.  The density time series from RAPID 26$^\circ$N  has been shifted forward in time by 7 months.  Positive lag corresponds to $26^\circ$N leading $16^\circ$N.}
\label{fignine}
\end{figure}

Lag correlations between density anomalies at $26^\circ$N and $16^\circ$N are computed for each depth (Fig.~\ref{fignine}a). Above 1200 m, there is little to no relationship between density anomalies at  $26^\circ$N and $16^\circ$N. Between 1200 m and 4650 m, density anomalies are correlated, with anomalies at $16^\circ$N tending to occur simultaneously or after those at $26^\circ$N.  Highest correlations are for RAPID leading MOVE by less than 1 year, with a secondary peak around 24 months (also with RAPID leading MOVE).  Strongest correlations occur around 1500 bar and 3500--4000 bar.  Fig.~\ref{fignine}b shows an example of a time series of density anomalies at 3800 dbar from RAPID $26^\circ$N  and MOVE $16^\circ$N, with the time series from $26^\circ$N  shifted forward by 7 months. Both latitudes show a shift from relatively dense to relatively light waters, at the end of 2009 at $16^\circ$N (and 7 months earlier at $26^\circ$N). These shifts are of the same sign and similar magnitude at the two latitudes. In the absence of compensating density changes on the eastern boundary, density anomalies at the west would represent large-scale coherent changes in transport in geostrophic shear in the subtropical North Atlantic, which we explore in the next section.

\section{Dynamic height and transports}

Transports from boundary arrays are calculated both from current meter measurements very near the west and from dynamic height differences between the west and east. Several previous investigations have separated the transport anomalies due to changes at the west from those in the east, to better identify the dynamic cause of those changes (See, for example, \citet{Kanzow-etal-2010}, \citet{Duchez-etal-2014}, and \citet{Elipot-etal-2014}). Here, we focus on the western boundary variations only, in order to compare like-with-like and because they have the greatest influence on low-frequency, deep transport variability.  We neglect transport fluctuations below 4820 dbar, including any northward flowing Antarctic Bottom Water.  These transports are expected to be small ($\sim$1 Sv at $26^\circ$N) with small variations (standard deviation of 0.4 Sv over 6 months) \citep{FrajkaWilliams-etal-2011,McCarthy-etal-2015}.

At $26^\circ$N, the geostrophic transport-per-unit-depth between the Bahamas and Canary Islands is derived from the thermal wind relation as
\begin{equation}
T_{int}(p)=\frac{\Phi_{east}(p)-\Phi_{west}(p)}{f}
\end{equation}
where $f$  is the Coriolis parameter, and $\Phi_{east}$ and $\Phi_{west}$ the dynamic height anomalies relative to zero at 4820 dbar at the east and west of the Atlantic, respectively. Full details of the calculation can be found in \citet{McCarthy-etal-2015}. Dynamic height is estimated from measured density profiles as
\begin{equation}
\Phi(p)=\frac{1}{\rho_0}\int_{4820}^p\delta(p')\,\mathrm{d}p'
\end{equation}
where $\rho_0$ is a constant reference density, and $\delta$ the specific volume anomaly ($1/\rho$). A different choice of reference level at $26^\circ$N can result in a different vertical structure of transports \citep{Roberts-etal-2013,Sinha-etal-2017}. At $16^\circ$N, numerical simulations suggested that the overturning transports could be recovered using dynamic height anomalies referenced to zero at depth \citep{Kanzow-etal-2008}.

\subsection{Dynamic height changes}
In considering western boundary variations only, we do not directly address the issue of the choice of reference level. Instead, we consider dynamic height anomalies relative to a deep reference level: 4820 dbar. Referenced to 4820 dbar, there is a clear shift from relatively low dynamic height anomalies to relatively higher dynamic height anomalies at both latitudes (Fig. 10). 
Similar to the time series of density anomalies (Fig. 9), this transition occurs around 2009 at $26^\circ$N, and some months later at $16^\circ$N. By construction, dynamic height anomalies are zero at 4820 dbar, and increasing upwards, with a further increase in the amplitude of anomalies across the depth range 2000--3500 dbar. Since dynamic height anomalies are constructed as a depth integral of inverse density anomalies (2), the density anomalies in the range 2000--3500 dbar are responsible for the changes in shear in transports through equation (1).   In comparison, dynamic height anomalies at the eastern boundary of $26^\circ$N (off the Canary islands) show markedly weak interannual variability (Fig. 11a), indicating little to no change in shear.  At the western flank of the Mid Atlantic Ridge at $16^\circ$N, the fluctuations still weaker than the western boundary of $16^\circ$N, indicating that even between the western boundary and the Mid Atlantic Ridge, shear changes are more strongly governed by changes at the western boundary.

\begin{figure}[h]
\centering
\includegraphics[width=20pc]{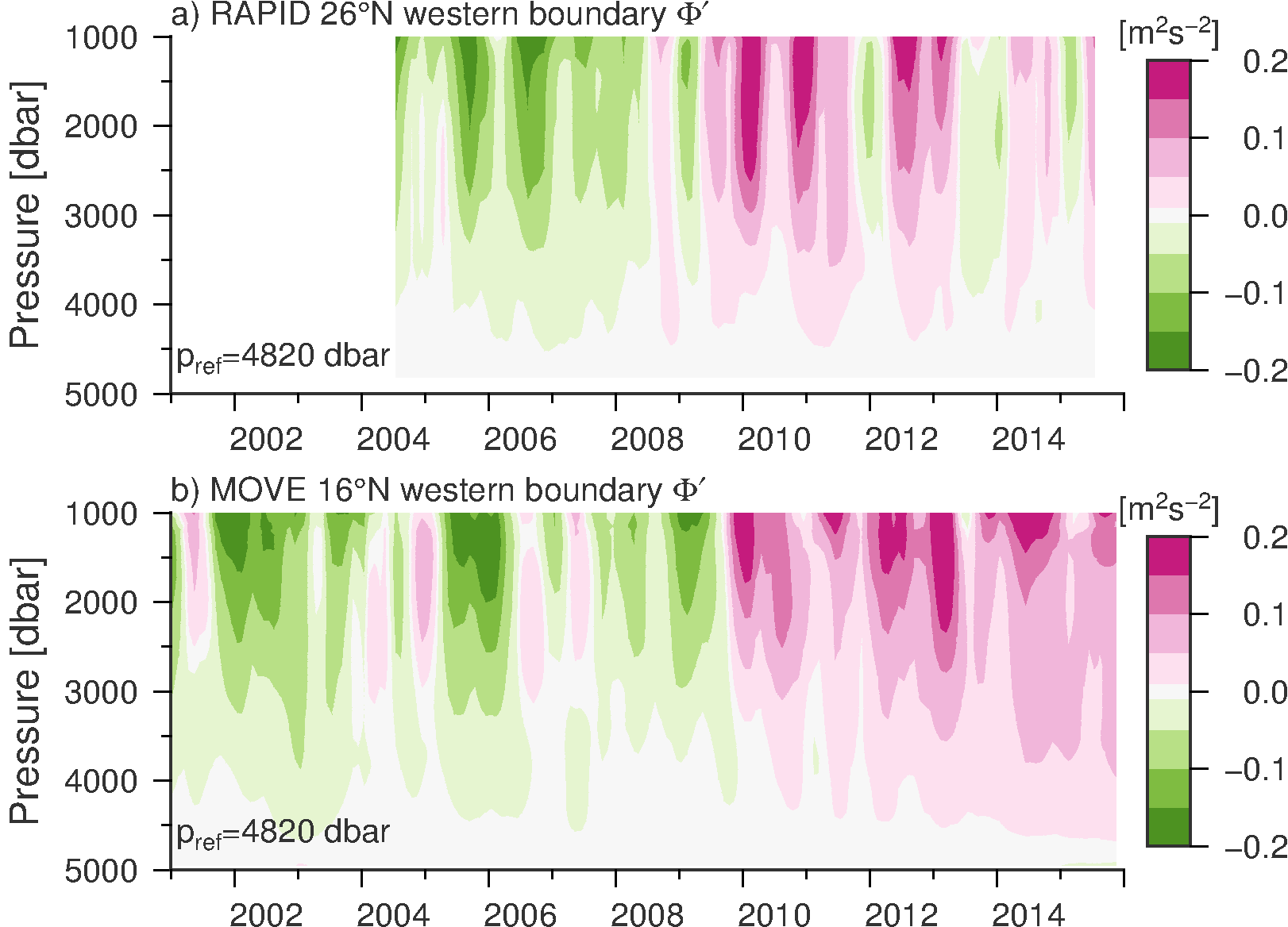}
\caption{Dynamic height anomalies at the western boundary of (a) RAPID $26^\circ$N and (b) MOVE $16^\circ$N, referenced to zero at 4820 dbar.  A transition from negative (green) to positive (pink) dynamic height anomaly at 1000 dbar indicates a relatively strengthening of the southward upper NADW (1000--3000 dbar) relative to the southward lower NADW (3000--5000 dbar).}
\label{figten}
\end{figure}

\begin{figure}[h]
\centering
\includegraphics[width=20pc]{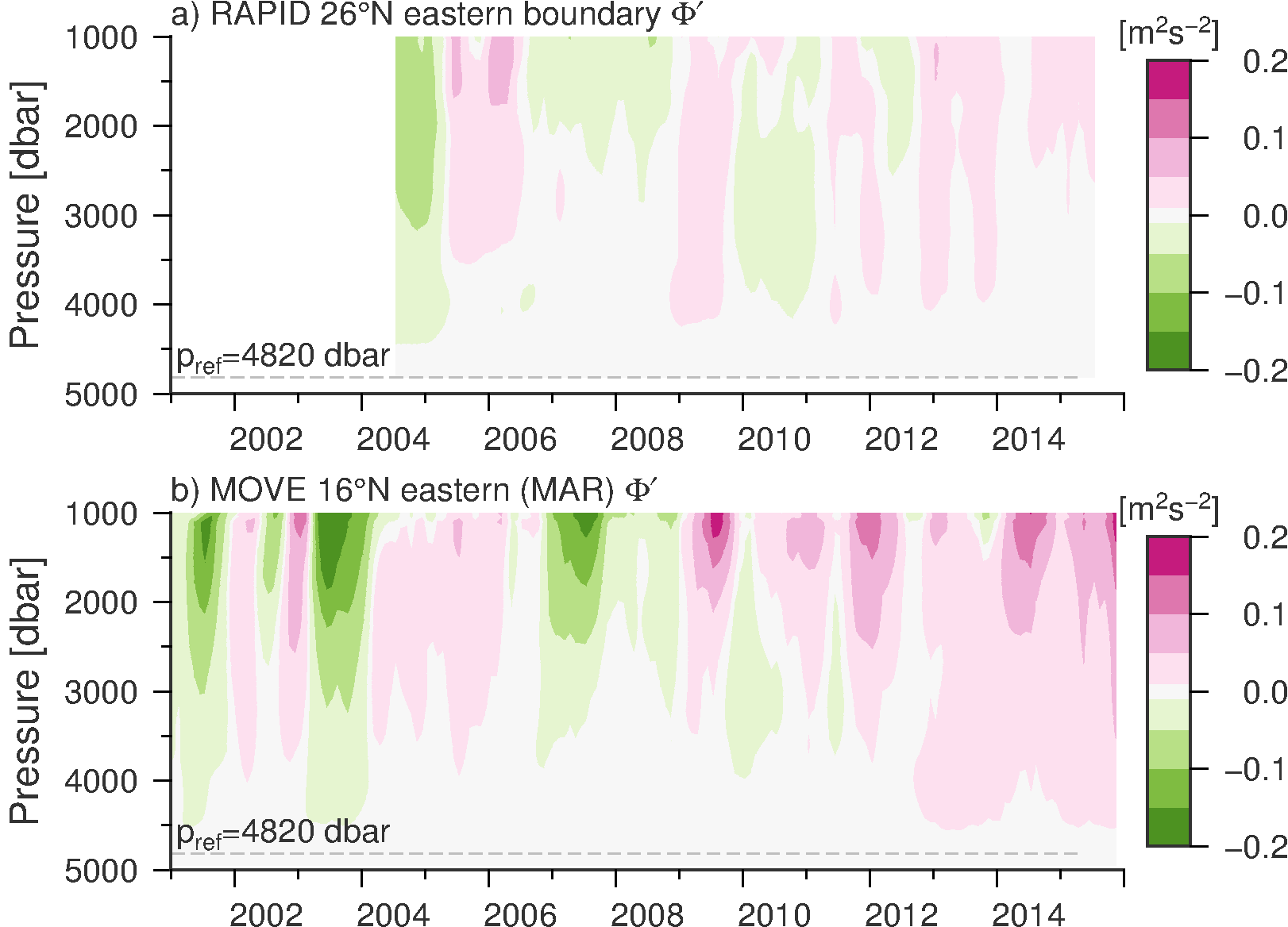}
\caption{As for Fig. 10, but for (a) the eastern boundary profile from RAPID $26^\circ$N and (b) the eastern mooring of MOVE $16^\circ$N (mooring MOVE1, west of the Mid-Atlantic Ridge).}
\label{figeleven}
\end{figure}

To quantify the fluctuations in shear at both latitudes---independent of the choice of reference level---we calculate the dynamic or geopotential thickness between two depths (Fig. 12). Over the 15 years of observations at $16^\circ$N, the thickness anomaly has shifted from negative to positive. At $26^\circ$N, the shift is the of similar magnitude but only documented over the past 10 years. The dynamic thickness anomaly is also calculated at the east (off of Africa for $26^\circ$N and at the western flank of the MAR for $16^\circ$N, see Fig. 12b). At $26^\circ$N, the deep dynamic height variability is negligible in the east. At $16^\circ$N, there are a few larger variations in 2007 and 2009, but for the most part, since 2004, the eastern boundary dynamic thickness anomaly has been relatively small.  Comparing the two 5-year periods (2004--2009 and 2009--2014), the dynamic thickness of this layer at $16^\circ$N changed from $12.22\pm0.07$ to $12.37\pm0.08$ $\mbox{m}^2\mbox{s}^{-2}$.  At $26^\circ$N, the change was from $12.32\pm0.05$ to $12.47\pm0.07$ $\mbox{m}^2\mbox{s}^{-2}$.  At both latitudes, the geopotential thickness change or shear increased by about 0.15 $\mbox{m}^2\mbox{s}^{-2}$. 

\begin{figure}[h]
\centering
\includegraphics[width=15pc]{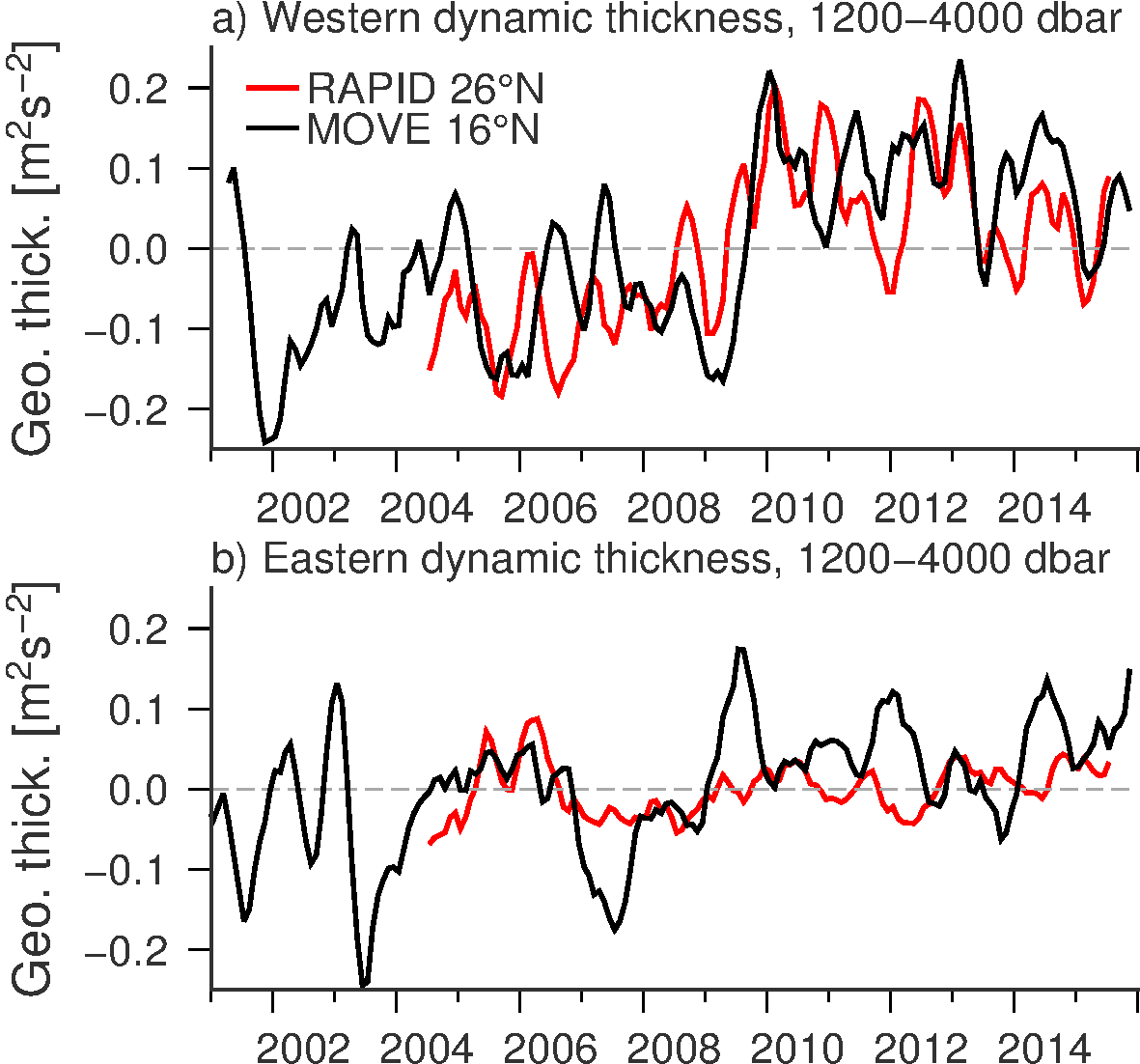}
\caption{Dynamic thickness anomaly time series at (a) the western boundary of RAPID $26^\circ$N and MOVE $16^\circ$N and (b) the eastern boundary of RAPID and the eastern mooring of MOVE (west of the Mid-Atlantic Ridge).}
\label{figtwelve}
\end{figure}

\begin{figure}[h]
\centering
\includegraphics[width=17pc]{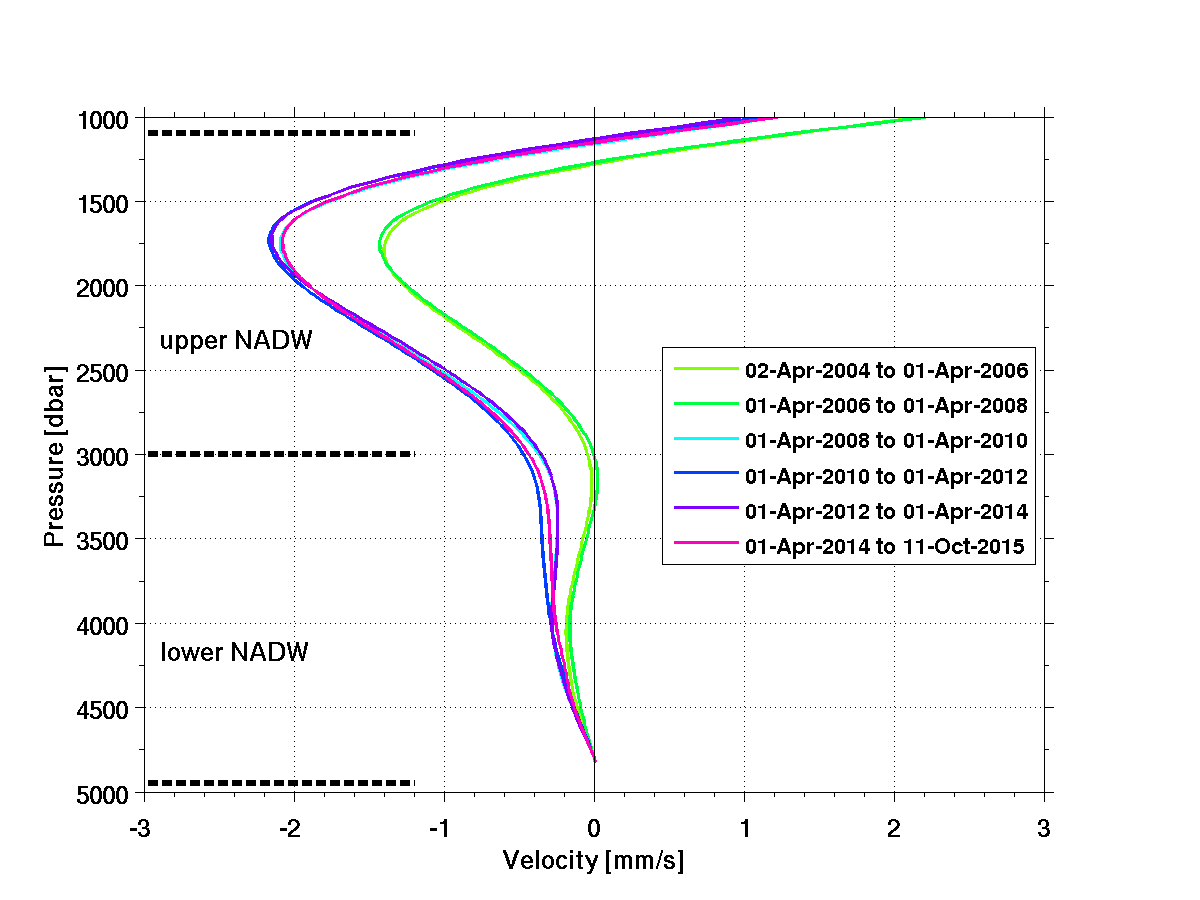}\includegraphics[width=17pc]{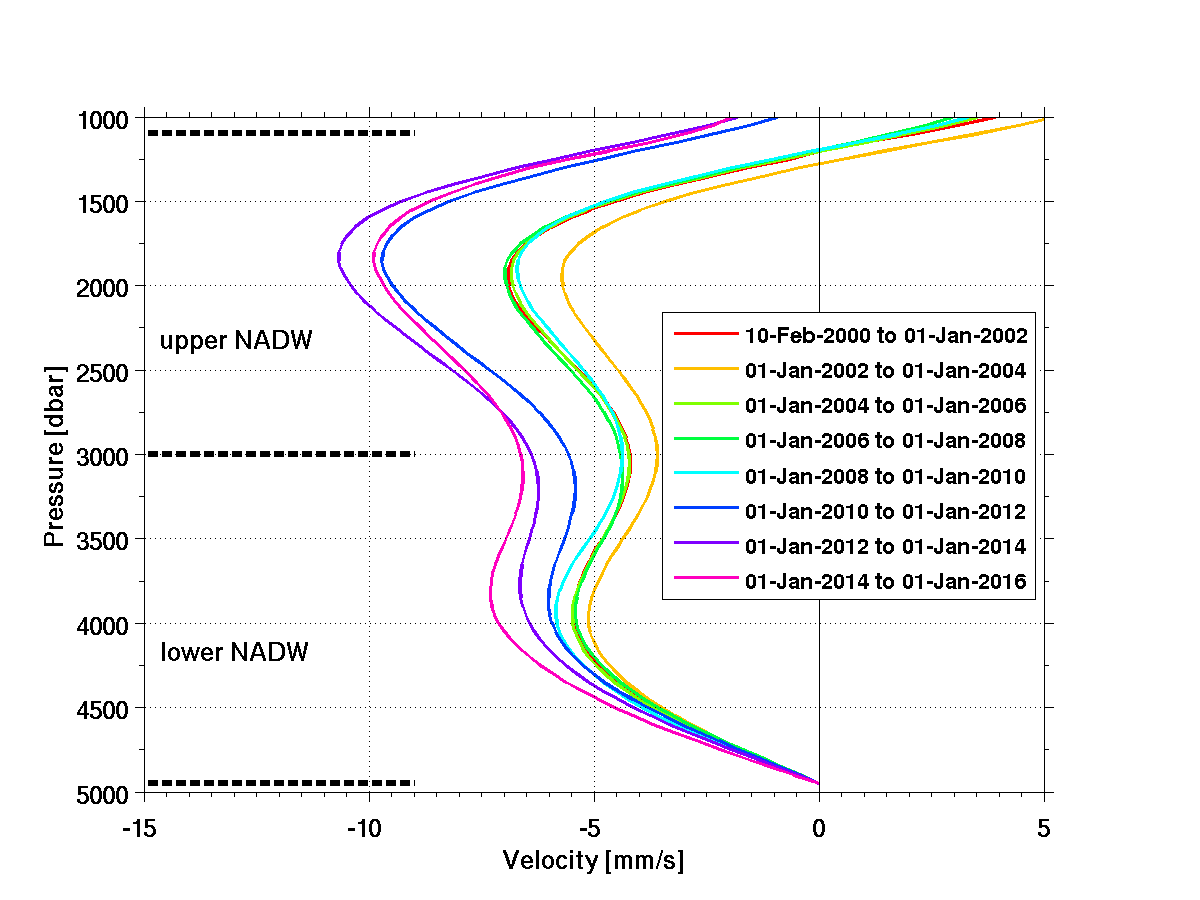}
\caption{Velocity estimates derived from dynamic height anomalies calculated at the western boundary profiles from (a)  RAPID $26^\circ$N and (b) MOVE $16^\circ$N, following \citep{Send-etal-2011}.  Dynamic height anomalies were integrated relative to a deep reference level.}
\label{figthirteen}
\end{figure}

\begin{figure}[h]
\centering
\includegraphics[width=15pc]{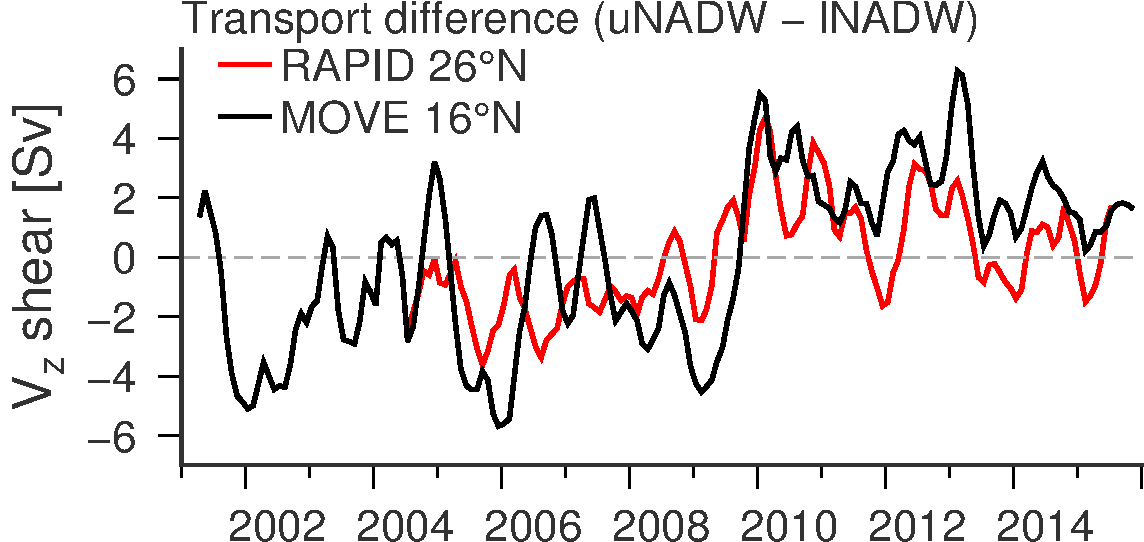}
\caption{Shear anomaly due to western boundary dynamic height changes as in (\ref{shear}) for MOVE at $16^\circ$N (black) and RAPID at $26^\circ$N (red). The observed dynamic height anomalies represent an increase in $\mathbf{V}_z$ by about 2.4 Sv (MOVE) and 2.7 Sv (RAPID) between the two 5-year periods, 2004--2009 and 2009--2014.}
\label{figfourteen}
\end{figure}

To put the dynamic height changes back into more relatable physical quantities, we can estimate the velocity profile due to dynamic height variations at the west only (Fig.~\ref{figfourteen}).  Here  we can clearly see that with a deep reference level, the change from earlier time periods to more recent periods is accompanied by a relative strengthening of the southward flow in the upper layer of transports (upper NADW, 1100--3000 dbar).  
Shear in the transport due to dynamic height anomalies at the west ($\Phi'_{west}$) can also be estimated between the two layers (1100--3000 m and 3000--5000 m) by scaling by $f$ and integrating in depth as
\begin{equation}
\mathbf{V}_z=\int_{3000}^{1100}\frac{-\Phi'_{west}(p)}{f}\,\mathrm{d}p - \int_{5000}^{3000} \frac{-\Phi'_{west}(p)}{f}\,\mathrm{d}p\label{shear}
\end{equation}
where the first integral represents the transport contribution from the intermediate layer (upper NADW), and the second integral from the lower layer (lower NADW).   Computing $\mathbf{V}_z$ at both latitudes gives a sense of the change of the circulation in units of Sv, where a positive value represents a strengthening of the upper NADW transports relative to the lower NADW transports (Fig.~\ref{figthirteen}).  Between the two five year periods, both latitudes showed an increase in the shear transport of 3.9~Sv (MOVE $16^\circ$N) and 2.7~Sv (RAPID $26^\circ$N).  Note that while the geopotential thickness anomaly at the two latitudes was similar, $f$ is smaller at $16^\circ$N resulting in a larger transport anomaly.

These results show that the observed density changes at the western boundary of the Atlantic are consistent in tendency (towards lighter water below 1200 dbar) and timing (between 2009--2010) at both latitudes.  This results in the same sign effect on changes to the geostrophic transport, $T_{int}$, relative to a deep level of no motion.  The effect of these changes is to intensify the shear between the lower and upper NADW layers (Fig.~\ref{figthirteen} \& \ref{figfourteen}).  At both latitudes, there is a notable change in the shear between these two layers between the earlier and latter part of the transport observations.

\subsection{Relationship between dynamic height and the MOC}

At $16^\circ$N, the strengthening of the upper NADW layer has been associated with an intensification of the deep southward flowing limb of the MOC.   With a reference level of zero velocity at 5000 m, the lower layer transports (3000--5000 m) are relatively constant over the 15 years, while the shear results in an intensification of the southward flow in the upper NADW layer (1100--3000 m, Fig.~\ref{figthirteen}). 

\begin{figure}[h]
\centering
\includegraphics[width=15pc]{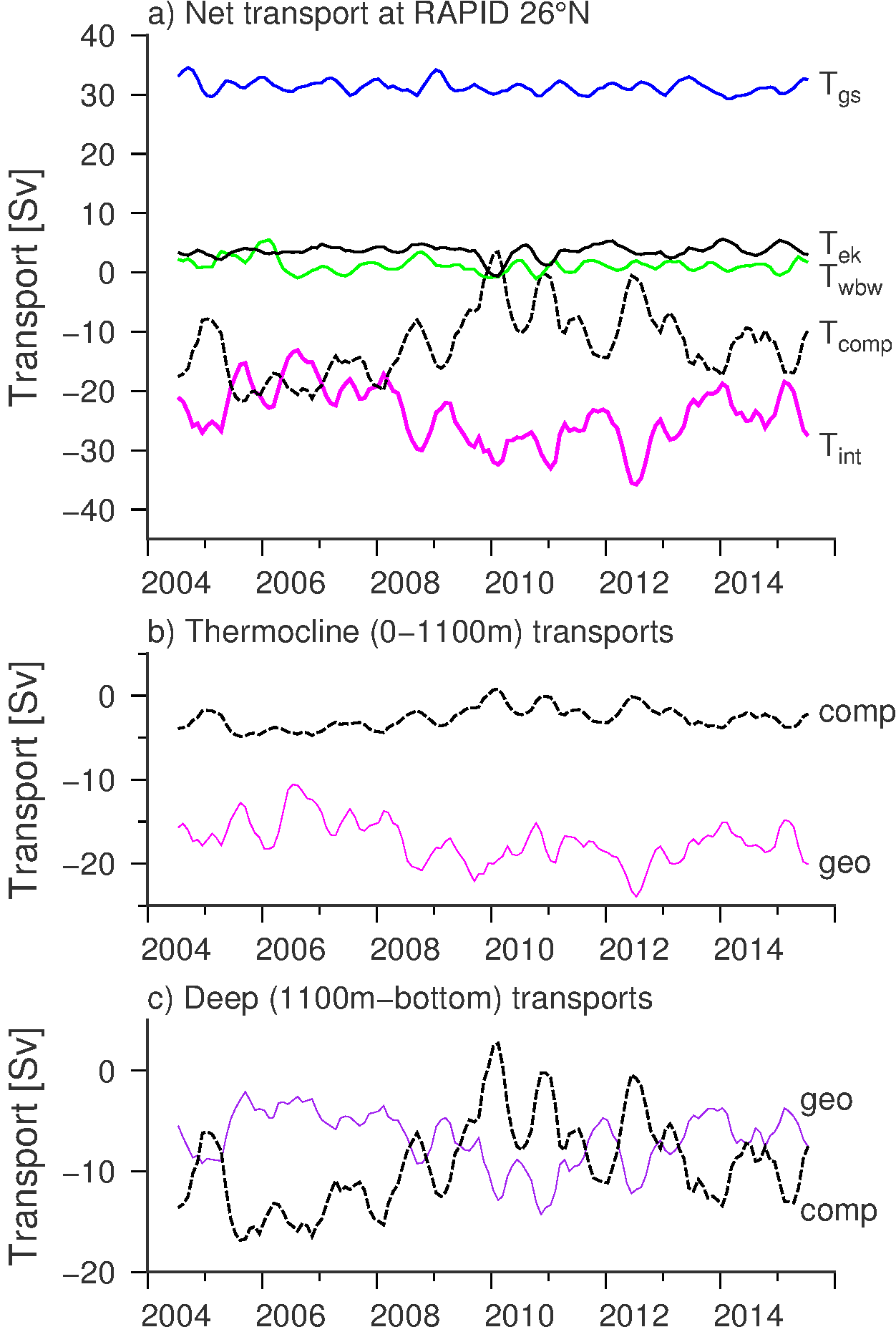}
\caption{Total transports at $26^\circ$N, applying mass compensation.  (a) Florida Current (blue), Ekman (black), western boundary current meter estimates (green), geostrophic estimates, relative to 0 at 4820 dbar (magenta) and external or compensation transport (black dashed).   The sum of these is zero at all times.  (b) Geostrophic transport in the thermocline (0--1100 m, magenta) relative to 0 at 4820 dbar, and the compensation applied over the 0--1100 m layer (black dashed).  (c) Geostrophic transport in the deep layer (1100m--bottom) and the compensation applied over this layer (black dashed).}
\label{figfifteen}
\end{figure}

At $26^\circ$N, however, the transports are derived from the geostrophic interior flow ($T_{int}$) as well as the compensation term, so that the overturning ($\Psi$) includes contributions from multiple components as
\begin{equation}
\Psi=\int^zT_{gs}+T_{ek}+T_{wbw}+T_{int}+T_{ext}\,\mathrm{d}z
\end{equation}
where $T_{gs}$ and $T_{ek}$ are the transports of the Florida Current and surface Ekman transport, respectively, and $T_{wbw}$ is from direct current meter observations in the western wedge.  $T_{gs}$ and $T_{ek}$ have little interannual variability over the 2004--2015 period (Fig.~\ref{figfifteen}a).  The change in dynamic height anomaly at the west results in an intensification of the southward interior flow ($T_{int}$) in the deep NADW layer (Fig.~\ref{figfifteen}c), while the thermocline shows a similar intensification of southward flow (Fig.~\ref{figfifteen}b).  These are similar to the changes estimated at $16^\circ$N.

However, the RAPID method applies a constraint of zero net mass transport across the section at $26^\circ$N.  This allows the compensation velocity (akin to the reference level velocity at depth) to change in time, and is encapsulated in the $T_{ext}$ term.  Due to the intensification of southward flow over all layers with time, weak changes in other components ($T_{gs}$, $T_{ek}$, and $T_{wbw}$), the compensation term must have an intensification of a northward flow  (Fig.~\ref{figfifteen}a).  The compensation is constructed as a uniform northward velocity since otherwise it would have been measured by the boundary arrays.  This means that when $T_{ext}$ it is integrated over the thick NADW layer (1100--5000 dbar) it has a larger contribution to the total transport in that layer than in the thermocline layer (0--1100 dbar).   The result is that the strengthening northward flow indicated by the $T_{ext}$ dominates over the strengthening southward flow of the $T_{int}$ in the deep layer, so that there is an overall weakening of the overturning circulation (southward flow below 1100 dbar).  The southward flow in the thermocline is still dominated the increasing southward flow of the $T_{int}$, and so over the 11 year record, there is a relative intensification of the southward flow in the thermocline in spite of a northward tendency in the $T_{ext}$.  This explains the origin of the opposing tendencies in the MOC estimated at $26^\circ$N, when compared to transports calculated at $26^\circ$N using a fixed, deep level of no motion.

\section{Conclusions}
Observations of the large-scale circulation at individual latitudes have been revolutionary to our understanding of variations in the overturning circulation \citep{Srokosz-Bryden-2015}. However, efforts to relate the variability observed at different latitudes via different measurement designs have proved challenging \citep{Elipot-etal-2013,Mielke-etal-2013,Elipot-etal-2014}. While the transport observations at $26^\circ$N and $16^\circ$N both rely on thermal wind (measuring density in order to calculate geostrophic shear), choosing an appropriate reference level to translate shear into velocity remains a challenge.  At $26^\circ$N, the reference level is applied as a barotropic compensation by assuming no net transport across the section on timescales longer than 10-days. At $16^\circ$N, the choice of no motion near 5000 m was determined by OSSEs and is consistent with bottom pressure measurements on sub-annual time scales.

Here, we have used the approach most consistent the RAPID and MOVE methodologies to investigate observed quantities (properties, density, dynamic height) which vary coherently or distinctly  between the two latitudes.  Overall, changes observed at the western boundary at $16^\circ$N and $26^\circ$N show consistent tendencies (towards freshening, lightening and an increase in deep shear of the southward flows), with similar magnitudes and a particular shift at both latitudes around 2009--2010.  The methods employed at these latitudes to determine the overturning differ from those used in other boundary arrays  (e.g., \citet{Toole-etal-2011} where the transport changes are identified in density space).   We find low frequency changes in density/dynamic height, occurring first at $26^\circ$N and less than a year later at $16^\circ$N.  These changes are computed directly from density observations, but may arise from either property changes (Fig. 2) or thickness/volume changes of a particular layer.  Due to complications with determining property changes on an isopycnal from a mooring with fixed point measurements, we leave that investigation for the future. 

This work highlights that the choice of reference level is a key element of estimating the overturning circulation through thermal wind balance.  At $26^\circ$N, application of a zero net mass transport constraint resulted in a reversal in the estimated deep transport tendencies (from a strengthening southward deep flow, consistent with a strengthening MOC to a weakening southward deep flow and reducing MOC).  It is possible that the calculated transports at the two latitudes, and their tendencies, are correct.  Estimates of the barotropic transport variability from PIES (Pressure inverted echo sounders) at $16^\circ$N support the use of a deep reference level, showing that even when incorporating deep pressure gradient fluctuations, the tendency of transport variability on timescales up to 2-years is not affected.  Due to limitations of measuring long records of pressure in the ocean, the barotropic flow cannot be evaluated over longer timescales.  The result of this analysis is that the baroclinic changes driven by the western boundary densities are consistent between the two latitudes; it is possible that variability in the barotropic transport  is distinct at the two latitudes.
  Assuming that the MOC at $16^\circ$N is  strengthening while the transport at $26^\circ$N is weakening, one could envision several ways for this to occur, including through changes to the largely unmeasured Antarctic Bottom Water flow or deep volume storage/release.  

It is beyond the scope of this paper to evaluate the method for determining the reference level at individual latitudes, or whether the practice of estimating overturning from dynamic height mooring arrays needs to be adjusted to eliminate or incorporate the uncertainties due to reference level choices. New investigations into using satellite-based estimates of ocean bottom pressure \citep{Bentel-etal-2015,Landerer-etal-2015} show promise at providing independent estimates of deep ocean transport variability but may also have trends in the data that are not related to circulation changes.  New investigations are underway to determine uncertainties in the RAPID $26^\circ$N method of calculating overturning transports \citep{Sinha-etal-2017}.  We conclude that of the observable robust changes over the 11- and 16-year moored observations, the southward flow in the subtropical North Atlantic below 3 km weakened relative to the southward flow above, and that a shift occurred between 2009--2010.  These results also highlight a critical area of uncertainty in estimating large-scale ocean transports from boundary measurements: that of how to best incorporate a choice of reference level in the geostrophic shear method.

\subsection*{Acknowledgments}
EFW was funded by a Leverhulme Trust Research Fellowship.  Data from the RAPID Climate Change (RAPID)/Meridional overturning circulation and heat flux array (MOCHA) projects are funded by the Natural Environment Research Council (NERC) and National Science Foundation (NSF, OCE1332978).  Data are freely available from www.rapid.ac.uk.    MOVE was also funded by National Oceanic and Atmospheric Administration (NOAA), the Climate Program Office--Climate Observation Division, and
initially by the German Bundesministerium fuer Bildung und Forschung.
MOVE is part of the international OceanSITES program (www.oceansites.org).   Florida Current transports are funded by the NOAA and are available from www.aoml.noaa.gov/phod/floridacurrent. Special thanks to the captains, crews, and technicians, who have been invaluable in the measurement of the MOC in the Atlantic over the past 15 years.






%
%
%
%
%
%
\bibliographystyle{apalike}


%
%
%







\end{document}